# Morphogenesis of Gastrovascular Canal Network in Aurelia Jellyfish: possible mechanisms

.

**SONG Solène[1], ZUKOWSKI Stanislaw[2,3], GAMBINI Camille[2], DANTAN Phillipe [2], MAUROY Benjamin[4], DOUADY Stéphane[2], CORNELISSEN Annemiek J.M. [2*]**

[1]LIS, UMR 7020, Turing Centre for Living Systems – Aix Marseille Univ – Université de Toulon – CNRS, Marseille, France

[2]Laboratoire MSC, UMR 7057, Université Paris Cité – CNRS, F-75006, Paris, France

[3]Institute of Theoretical Physics, Faculty of Physics, University of Warsaw, Warsaw, Poland

[4]Laboratoire J.A. Dieudonné, UMR 7351, Université Côte d'Azur – CNRS, Vader center, Parc Valrose, 06108 Nice Cedex 02, France

**\* Correspondence:**
annemiek.cornelissen@univ-paris-diderot.fr

**Keywords: morphogenesis, gastrovascular, network, jellyfish, instabilities, variability, mechanical constraints**

**Abstract**

Patterns in biology can be considered as predetermined, or arising from a self-organizing instability. Variability in the pattern can thus be interpreted as a trace of an instability, growing out from noise. Variability can thus hint toward an underlying morphogenetic mechanism. Here we present the variability of the gastrovascular system of the Jellyfish *Aurelia*. In this variability emerge a typical biased reconnection between canals, and correlated reconnections. Both phenomena can be interpreted as traces of mechanistic effects, the swimming contractions on the tissue surrounding the gastrovascular canals, and the mean fluid pressure inside them, respectively. This reveals the gastrovascular network as a model system to study morphogenesis of circulation networks and the morphogenetic mechanisms at play.

## 1 Introduction

Morphogenesis remains an important question in biology. Independently of how the phenotype can be selected through natural selection, it remains essential to understand how it can appear, develop from its original fertilized egg, and get its own shape. Since humans observe nature, they classify similar shapes into species. Within one species, the shape is robustly perpetuated across generations. So we know that biological shapes are constrained enough within one species. Even with the discovery of many genes and produced molecules, and their important role in morphogenesis, how these constraints are applied to guarantee a given result is not completely clarified. In particular how a complex shape can appear while being constrained remains obscure (1). A complex shape would need much information to be described, thus many regulations to achieve it. However, the unfolding in time of an instability can led to a regulated complex shape, from a simple mechanism(2).

In his pioneering work, Alan Turing proposed that, even if some morphogenesis can be implemented through the various concentrations of some chemical products, the pattern they present is created by a spatial instability (3). This means that even if the original distribution of chemicals is homogeneous, this state will be unstable and spontaneously goes into patches of different concentrations. And that the pattern is spontaneously created, not controlled: only its global characteristics, like wavelength, which depends on the basic characteristics of the chemicals, like reactions and diffusion characteristics, are imposed, not its particular position.

This view may seem contradictory with the constrained production of a stereotyped shape. However, many examples of fluctuating shapes can give the intuition of an underlying instability. This is the case we present here for the formation of the gastrovascular network of the jellyfish Aurelia aurita. Jellyfishes are very old life-forms that appeared before the "vertebrate" revolution, but still present a complex vascular structure. This vascular structure is an open circuit, perfusing the whole body from the open mouth to the stomach pouch and back (4, 5). The flow in this circuit is due both to the effect of the whole contraction of the body, and to the action of many cilia on the internal epithelium. These canals, from a basal life-shape, can be seen as an early simple model of a network of tubes with a transport function as the later evolved closed vascular networks.

This gastrovascular canal network develops while the jellyfish grows from its first ephyra stage. This ephyra stage emerges from a more primitive form of a polyp (5, 6) . During the transformation of the polyp into jellyfish (a process called strobilation), this polyp is subjected to an instability that creates many disks along its axial body, each disk being unstable in the radial direction and forming arms (lobes with two marginal lappets), of typical 8-fold symmetry. One after the other the top disk further transforms and eventually detaches, resulting in a free swimming jellyfish larva, the ephyra. This way of generation ensures that a series of jellyfish appears from a very same polyp, so they are clones. The development of the gastrovascular canal can be followed while the jellyfish goes from a starlike shape ephyra of a few mm size to a juvenile jellyfish of approximately 10 mm, which has just reached the circular shape of adult medusas, to a mature jellyfish of about 100mm.

We will focus on the growth of the pattern from juvenile, with few rather stereotyped canals, to adult, with many canals. We do not consider the formation of the first growth of connected canals, from the ephyra to juvenile jellyfish, which already have complex but more regular steps. This growth happens with the sprouting of new canals, and their reconnection with the rest of the system. There is a strong tendency of sprouting canals to reconnect to younger ones, leading in a perfect case to a particular fractal pattern. However, this bias is not absolute, and there are also many variations. With the observation of these dynamics, and their results of complex and varying shapes, we can get closer to the origin of the morphogenetic process. More precisely the question of which phenomena can be responsible for the development of these shapes can be discussed. In the following we will present two possible phenomena, which both have some interests and limitations.

## 2 Towards stereotypical canal network

### 2.1 Stereotypical Morphology

The gastrovascular canal network in juvenile jellyfish can be presented with a stereotypical structure (see Fig. 1A). For adult jellyfishes, it is tempting to present also a stereotypical shape, a fractal one, which can be sometimes observed (we found occurrences in nature or in Cherbourg Aquarium culture, see Materials and Methods). It can be described as canals connecting to each other, in a well-defined and precise order (see Fig. 1). To describe it the best is always, as for fractals, to come back to its construction, step by step. In $1/8^{th}$ of the jellyfish (an octant), there is radially one gastric pouch of the stomach (or the junction between two pouches) near the center, and a marginal ring canal, circling around the whole rim of the jellyfish. Radially, there are two canals, rather straight and unbranched, joining the pouches to the marginal ring canal, the adradial canals. Between such two straight canals, there is a canal joining the gastric pouch (interradial canal), or directly in the mouth opening at the pouch junction (perradial canal) to the marginal ring canal and a rhopalium (a sensory organ that can be caricatured as an "eye"), with two secondary side branches connecting the main inter- or perradial canal with the ring canal, forming a trifurcation. There is no apparent difference between adradial and perradial morphologies. We will thus simply call them "trifurcate" canals (in contrast we can call the adradial canals the "straight" canals). In juveniles, new canals mainly grow from the marginal ring canal, and connect to one of the surrounding already existing canals.

There is a strong tendency, around 80% of the cases (see supplementary), for the new sprouting canals to connect to the younger canal at proximity. In the juvenile state (fig. 1A) at the ring canal, the trifurcation cut the interval between the two straight canals in four. Roughly four new canals appear in these intervals, and connect either left or right to the two younger side canals of the fork. The next generation, of 8 canals, would also connect to the last previous generation, leading to a distinctive tree shape (fig 1B).

The essential question arising from this structure and its development is to understand why new canals would connect to the younger previous ones.

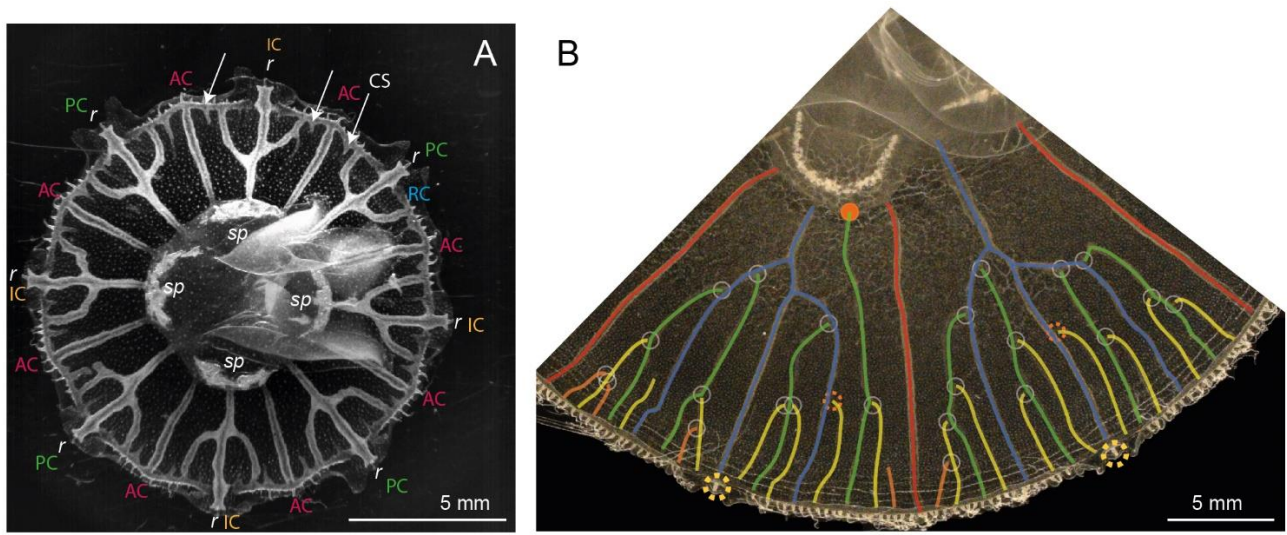

**Figure 1 A)** Picture of a juvenile jellyfish showing the structure of the gastro-vascular system. There are typically four stomach pouches (*sp*), and eight sensory organs, or rhopalium (*r*). At the periphery of the jellyfish there is a Ring Canal (RC, blue). Connecting this ring canal, from the sides of the stomach pouches, there are typically eight straight Adradial Canals (AC, red). Between these Adradial Canals, other canals connect either the stomach pouch, the Perradial Canal (PC, green), or the junction between two stomach pouches, the Interradial Canal (IC, orange), to the ropharia. New canals sprout (CS) from the Ring Canal (arrows).

**B)** Picture of two octants of a later developmental stage (original in Supplementary). The sprouting canals have reconnected to older ones, forming branched Perradial and Interradial Canal systems (rhopalium, yellow dashed circle, adradial canals, red, original trifurcate canals, blue). The sprouting canals have the tendency to reconnect to the youngest side canal (white circles), leading to a specific fractal tree shape. The four (green) canals sprouting in the two intervals between the fork and the two intervals near the side straight canals (red) would connect to side branches of the original (blue) fork . The next 8 canals (yellow) would connect to the previous one (green). And the next generation (orange) would connect to the previous one (yellow). Some connections do not follow this pattern, and either reconnect to older ones (dashed red circles), or even directly to the stomach pouch (red disk). Some irregular growth is also visible on the left (perraidal canals), leading to mixing of generations, as some fourth (orange) canals have already appeared, and even connected, while some third order (yellow) generations have not appeared yet or reconnected.

## 2.2 Differential contraction

To understand this differential connection of the sprouting canal to the youngest previous one, a first observation on the morphology and appearance of the canal itself is helpful. The canals, consisting of a canal wall with dense canal cells around a lumen, are situated inside a single largely spread cell sheet endoderm (see Fig. 2). We have observed that the cells around the tip of the sprout proliferate, forming a dense bilayer in a large pseudopod. The layers opens into two single layers of cells around a flat, smaller lumen (7, 8). When the canal sprout grows older, the lumen gets rounder and larger.

The second important point for a possible morphogenetic mechanism is that these canals are growing while the jellyfish is growing, and it is growing while being, since the beginning, actively contracting, to swim and gather food. These contractions from a nearly flat state to a bell shape is done by a muscle sheet contracting, and reducing the periphery perimeter. These movements then induce a considerable mechanical deformation to the endodermal layer, containing the canals, which is just above this muscle layer (Fig. 2).

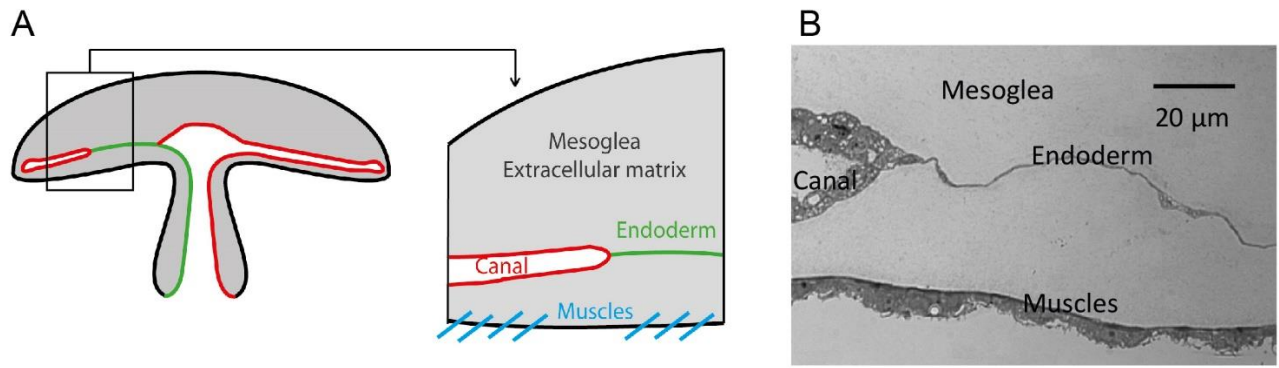

**Figure 2. A)** Schematic of position of the canals in the Jellyfish anatomy. The canals are inserted in an endodermal layer, near the lower epidermis with muscles. **B)** a microscopic cut showing the endodermal cell layer (left), connected to the canals cells (right) just above the epidermis layer and muscle cells (below).

A proposed mechanism for the differential connections is that the mechanical response to the contraction during swimming is different for different parts of the tissues. The endodermal cells which are not part of canals are submitted to a high mechanical stress as they are nearly incompressible, being held by the incompressible upper and lower mesoglea (9). On the contrary, canals are not flat, around a lumen, and the higher the cross section of the canal the more compressible they are (Fig. 3).

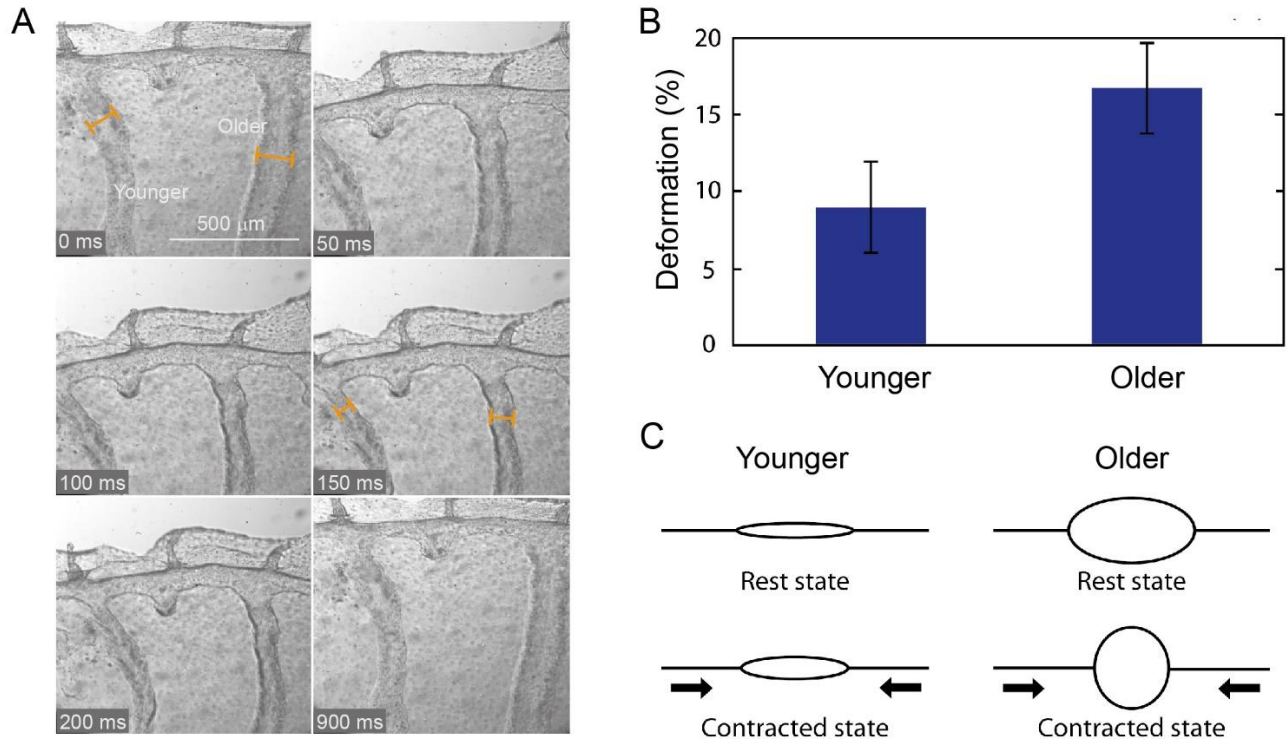

**Figure 3.** Relative contraction of canals during jellyfish swimming contraction. **A**: a sequence of pictures showing a young and an older canals pictured during a contraction (one can also observe a canal sprout). Canal width measurement during the resting phase and the maximum contraction allows to measure the relative contraction of the canals. **B**: the measurement shows that the contraction of a young canal is smaller than the one for older canals (9% compared to 17%). **C**: the schematic interpretation of the canals in resting and contracting state. Young canals are rather flat and cannot expand vertically during contraction. On the contrary, older canals are already open and this allows a larger contraction.

Numerical simulation (with 2D finite elements, see Material and Methods) of a contraction was thus performed, and suggests an accumulation of stress at the tip of a new canal (the stress is partly released at the new canal, and all the residual stress around focus on the tip), and the stress is different for the two surrounding canals of different age/stiffness. The quantitative result is that the maximum of stress is indeed shifted toward the younger stiffer canal (see Fig. 4)

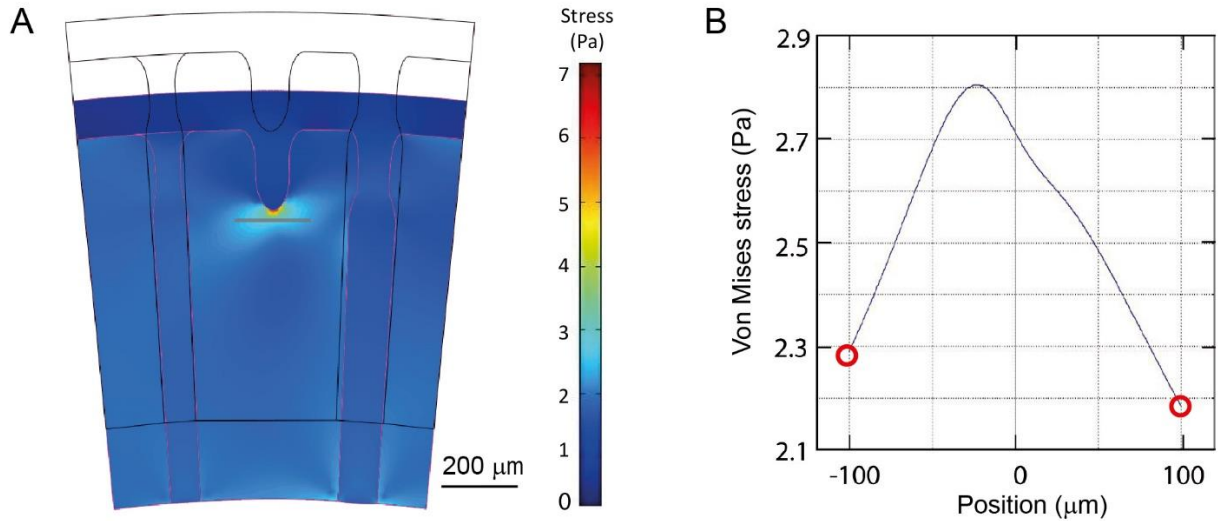

**Figure 4. A**: Finite element simulation with soft old canal (right) and stiff young canal (left). The contraction is reproduced by reducing the radial distance with a constant angle, thus reducing the orthoradial one. The mechanical stress (colored scale) is smaller (darker blue) in the soft canal, and reduced around the sprouting canal (dark blue), but concentrated at its tip (red spot), as in cracks. The value of the stress is plotted along a horizontal line just above the tip of the sprout (grey line). **B**: the measured value of the stress along the grey line. Its maximum is shifted toward the stiff young canal. This will turn the propagation of the new canal sprout toward the younger, softer one.

In this way canals can be seen as propagation of cracks in the endoderm: too high stress could induce the proliferation of cells and/or their transformation in canals cells, canals open and release the stress. The global stress field guides the movement of the tip of a new canal/crack.

## 3    Variability

Since the connection of the sprouting canal to a younger close one is only a strong bias, there are many variations of patterns, and only rarely a perfect one. Looking more generally at Aurelia jellyfish from different origins and growth conditions also reveal variable patterns.

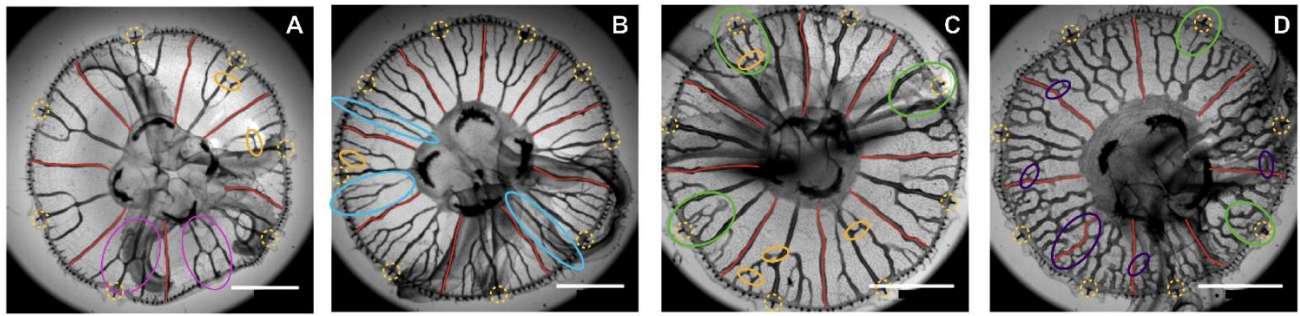

**Figure 5.** Four *Aurelia* specimens of around the same size, from Cherbourg, **A-D**, same images analyzed. The rhopalium are surrounded by yellow dashed circles, and the adradial canals drawn in red. In **A**, the continuation of canals after connecting to the central one is indicated by pink ellipses. In **A-B-C** some interconnection between canals are indicated (orange ellipses). In **B**, some side canals never reconnect to the central one, and connect directly to the stomach pouch (light blue ellipses). In **C** and **D**, some ropharia are not connected to canals (green ellipses). In **D**, there are many

oscillations and reconnections, making many loops, but in particular reconnection with straight adradial canals (violet ellipses). Scale bar 5 mm.

Looking at different jellyfishes of different origins, (from Cherbourg, which do not originate from the Roscoff strain, see Material and Methods), one can observe a great variety of patterns (Fig. 5). In this figure, one can observe that, even after reconnecting, the side canals would keep growing toward the stomach (pink ovals in Fig. 5-A). Another difference is the presence of oscillations along the canals that are potential sites for the growth of new canals. These growths induce the formation on large (old) jellies of loops, making the gastrovascular network looks locally like a foam (Fig. 5-D). This creates patterns that are much more difficult to analyze.

An interesting point is that since there is a variability in these patterns, one can study where these variations originate. In the case of ephyrae coming from a single polyp, which means they are clones, and grown together, thus in identical conditions, one can look at resulting patterns. Here we present three clones grown together. One can observe a similar type of pattern, but still differences (Fig. 6). This shows that, even for genetically identical jellyfish, and an identical environment, there is no strict control of the pattern. This suggests a self-organized pattern formation, relying on an instability, amplifying the noise.

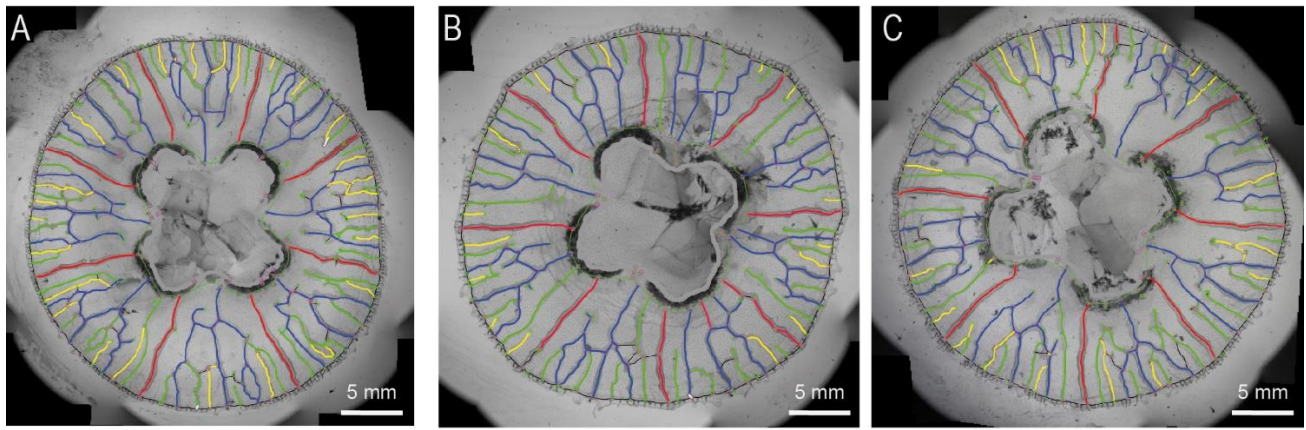

**Figure 6. A-B-C**: Three clones from the same polyp. The pattern has been interpreted for easier comparison. As in Fig.1B, the adradial canals are drawn in red; the central canal, and the first side canals forming the trifurcation are drawn in blue; the second generation in green; the third in yellow. One can observe many irregularities as in Fig. 5: side canals not reconnecting to the central one, reaching (or going to reach) the stomach pouch, interconnections making loops. But all these irregularities, although similar in the three clones, are in detail different.

## 4    Canal Breakthrough

### 4.1   Observation

A particular type of deviation from the stereotypical development, when one canal sprout connects directly to the stomach (Fig. 5-A), is interesting to see dynamically. During development, one can observe that some canal sprouts do not reconnect with the more central canals. Instead, they grow straight toward the stomach, independently from the other canals. It happens for the first generation canals that sprout between the fork and the side adradial canals (see Fig. 1B, 5B), around 17% of the time (see Supplementary). Interestingly, in such a case the next generation of side canals keep growing independently too (see Fig. 7) as long as the longest canal sprout did not reach the stomach pouch. However, as soon as the longest canal sprout reaches the stomach, it is observed that the smaller sprouts connect to the long canal that just got connected to the stomach.

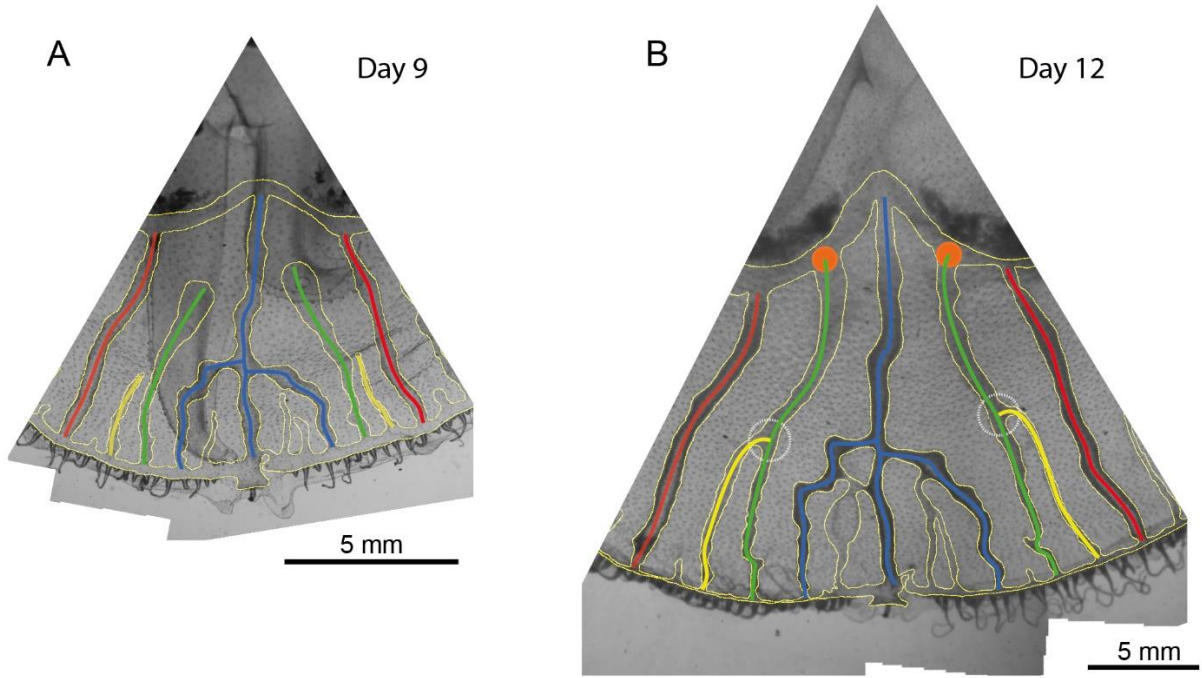

**Figure 7.** Two successive images of a jellyfish octant. **A**: at day 9, secondary (green) side canals didn't connect to the central ones, and are reaching toward the stomach pouches. Similarly, there are two younger (yellow) canals growing aside of them without connecting either. **B**: two days after, the side canals have reached the stomach pouch and connected (orange disks), and in the same period, the younger canals have also connected to these side canals (dashed white circles).

### 4.2 Interpretation

This coincidence of a canal breaking through the stomach pouch and the reconnection of a side canal to it, is reminiscent of reconnection observed in different physical systems. A typical morphogenetic instability, Saffman-Taylor fingering, driven by pressure gradient, is well known for creating irregular fractal branched structure, but without any reconnection between each finger. However, it was found that when resistance of the flow inside the fingers is taken into account, reconnection between a side canal and a longer one can occur in specific conditions (10).

Here this could happen if one considers that there is liquid pressure inside the growing canal sprouts. This could come from an asymmetrical average of the contractions in the radial direction, and to the movement of the cilia inside the already grown canals. When the canal sprouts are growing, they still have a large pressure. But when one canal sprout, in analogy with the Saffman-Taylor finger, reconnects, especially directly to the stomach, then its pressure drops to reach the pressure at the outlet. This dropping of pressure happens all along the canal, so that the lateral canal sprout can now perceive a place on the side with low pressure, and be attracted to it.

# 5     Discussion

When a pattern is constant, it is difficult to describe its origin and what controls it. On the contrary, variability helps to get closer to the mechanisms producing these patterns. Here we see that the pattern can be very variable, even in clonal jellyfish. That points to instabilities being at the origin of the pattern. Since Turing, we understand that instabilities mean that a homogeneous state is unstable, so that tiny inhomogeneities will grow to create a pattern. In this sense an instability starts with the amplification of noise. This first step results in a noisy pattern, which is a source of variability.

Later in development, some global constraints can regularize the pattern, as long-range interaction, leading to a periodic pattern. This is how regular and reproducible patterns can appear, even originating from an initial instability. But if the system keeps growing and being unstable, then this noise amplified variability can persist.

Here it seems that the gastro-vascular network can, under some growth conditions, follow a typical asymmetry (bias) and converge toward a stereotypical pattern. We could guess that not following the bias comes from the presence of more noise, for instance on the distance between canals, canal growth, and canal stiffness or resistivity. Such noise is well visible on Fig. 1B, left, where the generations of canals are irregular. This noise could blur the asymmetry and sometimes reverse it, leading to a non stereotypical connection.

The sprouting of new canals from the circular ring canal reveals also an instability of this ring, that would be similar to the oscillations of other parts of canals, leading to local sprouting and later to other reconnections, forming loops.

The fact that canals reconnect to each other is a particularly interesting phenomena. The gastro-vascular network is a tree structure connected to a ring canal. One would first imagine that it forms as a tree expanding with successive dichotomies of tips or side branching, and finally connecting to the ring canal. Here, we see a reverse growth: the branches appear from the ring canal, separated from each other, and reconnect only later. The reconnection between canals leads to the formation of loops. The sprouting from other canals than the ring canal, that further connect, form even more loops. This is interesting since the usual branching formation of trees, as in Laplacian growth, often forbids reconnection, hence the formation of loops (11). Such reconnection is thus a particular phenomena that deserves exploration.

The first mechanism proposed here relates to cracks, which are known to reconnect, being related to two dimensional stress (12)This relation to stress also explains the observed bias, that the crack is attracted to the larger stress, thus to the still stiff younger canal.

The second mechanism, even if related to Safmann Taylor and Laplacian growth, would happen in the special case of resistive fingers (10), that allows connection of side fingers when the pressure in the longer one is lower, for instance when it suddenly drops because of a breakthrough.

These two mechanisms could be happening in the jellyfish, or just be mechanical analogies of other phenomenons. But even on the mechanical point of view they are not incompatible, being driven by the stress in the endothelial layer and the pressure in the canals, which are complementary parts of the network.

The source of variability of the patterns, as shown in Fig. 5, should also be investigated. Is it due to different growth conditions, growth histories, or also to different strains, revealing a different sensibility to mechanical constraints for instance?

Globally, these first observations show that the gastrovascular network results from a spontaneous organization, or, in other words, that it appears from instabilities, enhancing noise, so that two growths never produce the same result even with settings as close as possible (clones from a single polyp grown together in the same conditions). There are clues that the morphogenesis of the pattern could be related to mechanical processes, since it grows while the jelly is swimming, with oscillating contractions. These contractions have clearly a mechanical effect on the tissue, either by direct contraction, or by secondary effect on the flow inside the already existing canals.

## 6 Material and Methods

### 6.1 Jellyfish culture

Jellyfish A. aurita were reared in the laboratory, at room temperature, in artificial seawater, produced by diluting 35 g or 28 g of synthetic sea salt (Instant Ocean; Spectrum Brands, Madison, WI) per liter of osmosis water (osmolarity 1100 mOsm). Polyps of the Roscoff strain (13) were obtained by courtesy of Konstantin Khalturin from the Marine Genomics Unit, Okinawa Institute of Science and Technology Graduate University, Onna, Okinawa, Japan. Strobilation in polyps was induced by a lowering of temperature down to 10°C (14). The newborn ephyrae were bred to adult stage. The measurements were performed on jellyfish at different sizes of juvenile jellyfish. Juvenile jellyfish had just reached the circular shape of adult medusas with a diameter of ~1 cm. Juveniles grow out into adult jellyfishes with fully developed stomach pouches.

Juvenile jellyfish (~1 cm in diameter) were obtained from 'Jellyfish Concept' in Cherbourg from their culture. The original polyps are extracted from the North Sea around Cherbourg. Juvenile jellies were bred to adults while growing. In the manuscript we refer to these jellyfish as 'Cherbourg jellyfish', when they originate from the Roscoff strain we do not specify it in the manuscript.

### 6.2 Imaging of the gastrovascular canal network

The gastrovascular network of the jellyfish was observed using a Leica macro zoom (MACROFLUO LEICA Z16 APO S/No: 5763648) and a photron fastcam SA3 camera, or directly using a Nikon D3300 camera with macro lens AF-S DX Micro NIKKOR 40mm f/2.8G. Jellyfish were catch from the aquarium approximately 3 hours after feeding with artemia when the gastrovascular canals colored orange from the digested artemia. When they reach about 2.5 cm in diameter, jellyfish were anesthetized with magnesium chloride dissolved in water having the same salinity as the artificial seawater in which they are swimming. To anesthetize the jellies, the volume of the jellyfish with seawater was doubled with the magnesium chloride solution. Then, they are put in a petri dish in shallow seawater with the sub umbrella facing up. The images are taken by transillumination.

### 6.3 Numerical Simulation

Numerical simulations were performed with finite elements toolbox COMSOL Multiphysics 3.5a (15). We approximated the endoderm and the canals as 2D surface elements with different stiffness, for simplicity and speed of simulation.

The simulations were performed on a small piece of a ring at the edge of a circular disk with a radius of 5 mm, with a radial length of 1 mm and with a 12 degrees angle. The geometry of the canals and sprout were chosen to be coherent with our observations in a juvenile jellyfish of 1 cm diameter. The geometry with the simulation mesh is shown in the Supplementary figure S7 A.

We assume the endoderm as a flat rigid incompressible elastic sheet (Young's modulus $E_{endo} = 100$ Pa and Poisson's modulus Poisson's ratio $\nu_{endo} \sim 0.5$). The simulations in 2D imply the absence of out of plane buckling. This assumption is justified for small juveniles at the onset of contraction since the endoderm is held in plane by the mesoglea located above and below.

We modeled the canals in 2D by a slightly compressible elastic membrane, with lower Young's moduli than the endoderm. The young's modulus of the flat young canal is assumed to be stiffer with $E_{youngcanal} = 30$ Pa than the rounder old canal, $E_{oldcanal} = 10$ Pa. The Poisson's ratio of the canals equals 0.3 ($\nu_{canal} = 0.3$), which allows for compression.

We observed that by choosing the elastic modulus of the canals 10 times lower than the endoderm, we obtained rates of reduction of the diameter of the ducts close to those observed in vivo (Figure 2B). Note that if the value of stress in the endoderm changes with the value of the Young's moduli of the endoderm and the canals, the distribution of these stresses does not depend on the value of the Young modulus of the endoderm, as long as it remains higher than that of the canals.

To simulate the muscular contraction we impose a circular deformation by reducing the ring of the disk radially by 200 um in 1 sec, resulting in a gradually increasing circular deformation from 4% at the outer edge (top of ring canal), and 5 % at the inner edge. In radial direction, zero deformation was imposed. The deformation of the boundaries are shown in the Supplementary figure S7 B.

The different values of the Young's and Poisson's moduli of the different elements under compression results in a distribution of normal radial mechanical compressive stresses $\sigma_{r,r}$, normal circular stresses ($\sigma_{r\varphi,r\varphi}$) and shear stresses ($\sigma_{r,r\varphi}$) which are accumulated at the tip of the sprout.

The von Mises stresses, obtained by combining these different stresses(15), give a satisfactory scalar representation of the stress distribution in the endoderm.

# 7    Conflict of Interest

*The authors declare that the research was conducted in the absence of any commercial or financial relationships that could be construed as a potential conflict of interest.*

# 8    Author Contributions

SS performed experiments, analyses of the network, in particular for the clones, and wrote the paper. ZS performed experiments, analysed the growth dynamics, in particular for the breakthrough event, and wrote the paper. CG performed experiments, in particular for the anatomy, analyses, in particular the contractions, and modelled them mechanically, and performed mechanical numerical simulations. DP performed mechanical numerical simulations. MB defined the numerical simulations, and wrote the paper. DS analysed the patterns, and wrote the paper. CAJM directed the work, performed experiments, analyses, modelling, and wrote the paper.

## 9 Funding

This work was supported by The LABEX "WHO AM I?" (No.ANR-11-LABX-0071) with a PhD fellowship for Solène Song (doctorants 2015) and a collaborative grant (Projets collaboratifs 2013-II). And by the Mission for Transversal and Interdisciplinary Initiatives (MITI) of the French National Centre of Scientific Research (CNRS), AAP Auto-organisation 2021 and 2022.

## 10 Acknowledgments

We thank Carine Vias and Léna Zig for their excellent care for the jellyfish. We thank Vincent Fleury for pointing to us this biological system.

**12   Legends Figures**

**Figure 1 A)** Picture of a juvenile jellyfish showing the structure of the gastro-vascular system. There are typically four stomach pouches (*sp*), and eight sensory organs, or rhopalium (*r*). At the periphery of the jellyfish there is a Ring Canal (RC, blue). Connecting this ring canal, from the sides of the stomach pouches, there are typically eight straight Adradial Canals (AC, red). Between these Adradial Canals, other canals connect either the stomach pouch, the Perradial Canal (PC, green), or the junction between two stomach pouches, the Interradial Canal (IC, orange), to the ropharia. New canals sprout (CS) from the Ring Canal (arrows).

**B)** Picture of two octants of a later developmental stage (original in Supplementary). The sprouting canals have reconnected to older ones, forming branched Perradial and Interradial Canal systems (rhopalium, yellow dashed circle, adradial canals, red, original trifurcate canals, blue). The sprouting canals have the tendency to reconnect to the youngest side canal (white circles), leading to a specific fractal tree shape. The four (green) canals sprouting in the two intervals between the fork and the two intervals near the side straight canals (red) would connect to side branches of the original (blue) fork . The next 8 canals (yellow) would connect to the previous one (green). And the next generation (orange) would connect to the previous one (yellow). Some connections do not follow this pattern, and either reconnect to older ones (dashed red circles), or even directly to the stomach pouch (red disk). Some irregular growth is also visible on the left (perraidal canals), leading to mixing of generations, as some fourth (orange) canals have already appeared, and even connected, while some third order (yellow) generations have not appeared yet or reconnected.

**Figure 2.  A)** Schematic of position of the canals in the Jellyfish anatomy. The canals are inserted in an endodermal layer, near the lower epidermis with muscles. **B)** a microscopic cut showing the endodermal cell layer (left), connected to the canals cells (right) just above the epidermis layer and muscle cells (below).

**Figure 3.** Relative contraction of canals during jellyfish swimming contraction. **A**: a sequence of pictures showing a young and an older canals pictured during a contraction (one can also observe a canal sprout). Canal width measurement during the resting phase and the maximum contraction allows to measure the relative contraction of the canals. **B**: the measurement shows that the contraction of a young canal is smaller than the one for older canals (9% compared to 17%). **C**: the schematic interpretation of the canals in resting and contracting state. Young canals are rather flat and cannot expand vertically during contraction. On the contrary, older canals are already open and this allows a larger contraction.

**Figure 4. A**: Finite element simulation with soft old canal (right) and stiff young canal (left). The contraction is reproduced by reducing the radial distance with a constant angle, thus reducing the orthoradial one. The mechanical stress (colored scale) is smaller (darker blue) in the soft canal, and reduced around the sprouting canal (dark blue), but concentrated at its tip (red spot), as in cracks. The value of the stress is plotted along a horizontal line just above the tip of the sprout (grey line). **B**: the measured value of the stress along the grey line. Its maximum is shifted toward the stiff young canal. This will turn the propagation of the new canal sprout toward the younger, softer one.

**Figure 5.** Four *Aurelia* specimens of around the same size, from Cherbourg, **A-D**, same images analyzed. The rhopalium are surrounded by yellow dashed circles, and the adradial canals drawn in red. In **A**, the continuation of canals after connecting to the central one is indicated by pink ellipses. In **A-B-C** some interconnection between canals are indicated (orange ellipses). In **B**, some side canals never reconnect to the central one, and connect directly to the stomach pouch (light blue ellipses). In **C** and **D**, some ropharia are not connected to canals (green ellipses). In **D**, there are many oscillations and reconnections, making many loops, but in particular reconnection with straight adradial canals (violet ellipses). Scale bar 5 mm.

**Figure 6. A-B-C**: Three clones from the same polyp. The pattern has been interpreted for easier comparison. As in Fig.1B, the adradial canals are drawn in red; the central canal, and the first side canals forming the trifurcation are drawn in blue; the second generation in green; the third in yellow. One can observe many irregularities as in Fig. 5: side canals not reconnecting to the central one, reaching (or going to reach) the stomach pouch, interconnections making loops. But all these irregularities, although similar in the three clones, are in detail different.

**Figure 7.** Two successive images of a jellyfish octant. **A**: at day 9, secondary (green) side canals didn't connect to the central ones, and are reaching toward the stomach pouches. Similarly, there are two younger (yellow) canals growing aside of them without connecting either. **B**: two days after, the side canals have reached the stomach pouch and connected (orange disks), and in the same period, the younger canals have also connected to these side canals (dashed white circles).